\documentclass[reprint,superscriptaddress,
 amsmath,amssymb,
 aps,
 prb,
]{revtex4-1}

\usepackage{graphicx}
\usepackage{dcolumn}
\usepackage{bm}
\usepackage{natbib}

\begin{document}

\preprint{APS/123-QED}

\title{Interplay between localized and itinerant magnetism in Co substituted FeGa$_{3}$
}

\author{A.~A.~Gippius}
\affiliation {Department of Physics, Moscow State University, 119991, Moscow, Russia}
\affiliation{A.V. Shubnikov Institute of Crystallography, 119333, Moscow, Russia}

\author{V.~Yu.~Verchenko}
\affiliation{Department of Chemistry, Moscow State University, 119991, Moscow, Russia}
\affiliation{National Institute of Chemical Physics and Biophysics, Tallinn, Estonia}

\author{A.~V.~Tkachev}
\affiliation {Department of Physics, Moscow State University, 119991, Moscow, Russia}
\affiliation{A.V. Shubnikov Institute of Crystallography, 119333, Moscow, Russia}

\author{N.~E.~Gervits}
\affiliation {A.V. Shubnikov Institute of Crystallography, 119333, Moscow, Russia}

\author{C.~S.~Lue}
\affiliation {Department of Physics, National Cheng Kung University, 70101, Tainan, Taiwan}

\author{A.~A.~Tsirlin}
\affiliation {National Institute of Chemical Physics and Biophysics, Tallinn, Estonia}

\author{N.~B\"uttgen }
\affiliation {Experimental Physics V, University of Augsburg, 86159, Augsburg, Germany}

\author{W.~Kr\"atschmer }
\affiliation {Experimental Physics V, University of Augsburg, 86159, Augsburg, Germany}

\author{M.~Baenitz }
\affiliation {Max Planck Institute for Chemical Physics of Solids, 01187 Dresden, Germany}

\author{M.~Shatruk }
\affiliation {Department of Chemistry and Biochemistry, Florida State University, Tallahasee, FL 32306, USA}

\author{A.~V.~Shevelkov }
\affiliation {Department of Chemistry, Moscow State University, 119991, Moscow, Russia}

\date{\today}

\begin{abstract}
The evolution of the electronic structure and magnetic properties with Co substitution for Fe in the solid solution Fe$_{1-x}$Co$_x$Ga$_3$ was studied by means of electrical resistivity, magnetization, \textit{ab initio} band structure calculations, and nuclear spin-lattice relaxation $1/T_1$ of the $^{69,71}$Ga nuclei. Temperature dependencies of the electrical resistivity reveal that the evolution from the semiconducting to the metallic state in the Fe$_{1-x}$Co$_x$Ga$_3$ system occurs at $0.025<x<0.075$. The $^{69,71}(1/T_1)$ was studied as a function of temperature in a wide temperature range of $2\!-\!300$\,K for the concentrations $x = 0.0$, 0.5, and 1.0. In the parent semiconducting compound FeGa$_3$, the temperature dependence of the $^{69}(1/T_1)$ exhibits a huge maximum at about $T\!\sim\!6$\,K indicating the existence of in-gap states. The opposite binary compound, CoGa$_3$, demonstrates a metallic Korringa behavior with $1/T_1$ $\propto T$. In Fe$_{0.5}$Co$_{0.5}$Ga$_3$, the relaxation is strongly enhanced due to spin fluctuations and follows $1/T_1\propto T^{1/2}$, which is a unique feature of weakly and nearly antiferromagnetic metals. This itinerant antiferromagnetic behavior contrasts with both magnetization measurements, showing localized magnetism with a relatively low effective moment of about 0.7\,$\mu_B$/f.u., and \textit{ab initio} band structure calculations, where a ferromagnetic state with an ordered moment of 0.5\,$\mu_B$/f.u. is predicted. The results are discussed in terms of the interplay between the localized and itinerant magnetism including in-gap states and spin fluctuations.

\end{abstract}

\pacs{Valid PACS appear here}
\maketitle


\section{Introduction}

Solid solutions based on FeGa$_3$ attracted much interest because of prospective thermoelectric applications and an intriguing low-temperature magnetic behavior. The parent binary compound FeGa$_3$ is a rare representative of non-magnetic and semiconducting Fe-based intermetallic compounds \cite{T1} akin to the small gap semiconductors FeSi\cite{T2,T3} and FeSb$_2$.\cite{T4,T5} Because of the small value of the energy gap ($E_g$) and narrow energy bands, these compounds are considered as potential thermoelectric materials demonstrating extremely high Seebeck coefficient values of $|S|\approx 500$~mV/K at 50~K and $|S|\approx 45$~mV/K at 10~K for FeSi\cite{T6} and FeSb$_2$,\cite{T7} respectively. The value of $E_g$ in FeGa$_3$ was determined by various theoretical and experimental techniques, including \textit{ab initio} band structure calculations ($E_g=0.4-0.5$~eV),\cite{T8,T9} high-temperature magnetometry ($E_g=0.3-0.5$~eV),\cite{T10} photoelectron spectroscopy ($E_g\leq 0.8$~eV),\cite{T10} and resistivity measurements ($E_g=0.14$~eV).\cite{T11} As discussed in Ref.~\onlinecite{T9}, the dominant contribution to the density of states (DOS) near the Fermi level in FeGa$_3$ comes from the Ga $4p$ states and from predominantly non-bonding Fe $3d$ states at the top of the valence band. The band gap originates from a strong hybridization of the Fe $d$ and Ga $p$ atomic orbitals. The formation of the energy gap in FeGa$_3$ is reminiscent of that in strongly correlated $3d$ and $4f$ Kondo-insulators including FeSi and FeSb$_2$, which are likewise characterized by a small hybridization gap at the Fermi level.\cite{T12,T13}

It has been reported\cite{T14} that already a few percent cobalt doping in FeGa$_3$ drastically changes the properties of the parent compound. Namely, the 5\% Co-doped Fe$_{0.95}$Co$_{0.05}$Ca$_3$ exhibits properties of a bad metal and a Curie-Weiss paramagnet, in contrast to semiconducting and nonmagnetic FeGa$_3$. Our \textit{ab initio} band structure calculations have shown that the Co doping shifts the Fermi level position towards the conduction band formed by the $3d$ (Fe/Co) and $4p$ (Ga) orbitals, thus leading to metallic properties and enabling precise tuning of the power factor PF = $S^2\sigma$ (Ref.~\onlinecite{T9}), where $\sigma$ is the electrical conductivity. According to the resistivity data, the true metallic state for the Fe$_{1-x}$Co$_x$Ga$_3$ solid solution is achieved when $x=0.125$.\cite{T9} The end member of the solid solution, CoGa$_3$, exhibits good metallic properties with residual resistivity ratio at the order of 100 and a temperature-independent paramagnetic susceptibility of conduction electrons, which is, however, outweighed by the core diamagnetic contribution.\cite{T1,T14}

Remarkably, the Ge doping on the Ga site has an even more dramatic effect on the magnetism.\cite{T15} The FeGa$_{3-y}$Ge$_y$ solid solution reveals ferromagnetic ordering and an associated quantum critical behavior close to the critical concentration of $y_c = 0.13$, where the ferromagnetism emerges. In contrast, the Fe$_{1-x}$Co$_x$Ga$_3$ solid solution remains paramagnetic or nearly paramagnetic for all Co concentration studied so far.\cite{T15}

\begin{figure}
\center{\includegraphics[width=1\linewidth]{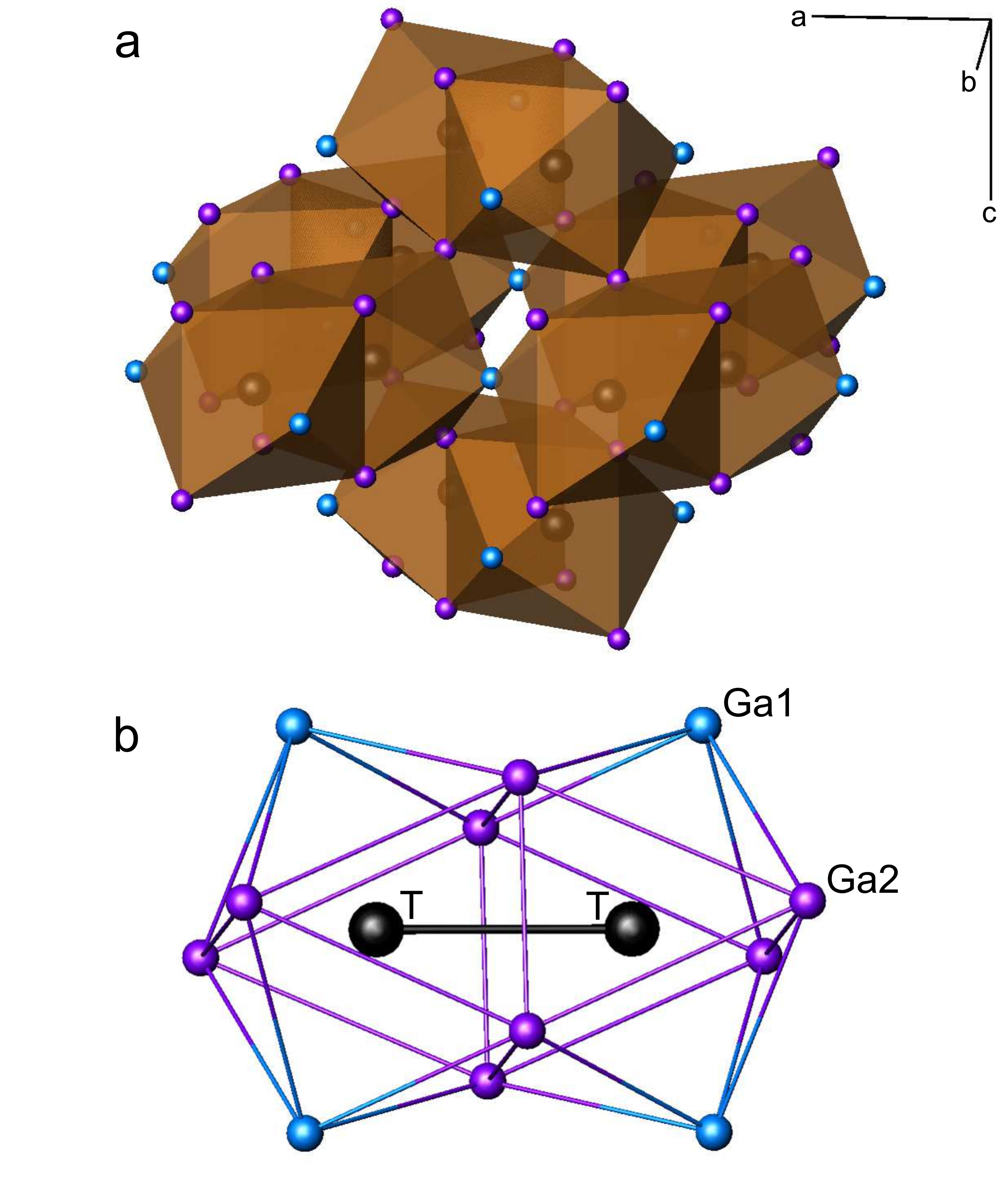}}
\caption{Polyhedral representation of the (a) crystal structure of the Fe$_{1-x}$Co$_x$Ga$_3$ solid solution, and (b) its building unit - a pair of bicapped trigonal prisms centered by the T--T dumbbell.}
\label{S1}
\end{figure}

The Fe$_{1-x}$Co$_{x}$Ga$_{3}$ solid solution exists for any Co concentration ($x$) and demonstrates a noticeable deviation from the Vegard behavior, even though the crystal structure and lattice symmetry remain the same for all Co concentrations.\cite{T9} This crystal structure should be considered as a three-dimensional framework constructed by polyhedra (Figure~\ref{S1}a). The main building unit is a pair of face-shared bicapped trigonal prisms centered by a T--T dumbbell, where T = Fe, Co (Figure~\ref{S1}b). The assembly of building units is arranged in compliance with the four-fold screw axis, which is parallel to the $c$ direction, such that face-shared filled prisms and empty spaces alternate in a staggered order. Thus, the assembly of polyhedra constructs the entire, almost isotropic crystal structure. In support of this, the isostructural compounds FeGa$_3$\cite{T16} and RuIn$_3$\cite{T17} show no significant anisotropy in their transport properties. On the other hand, electronic properties and their evolution upon doping may critically depend on the local structure, i.e., whether the \mbox{Fe--Fe}, Fe--Co, or Co--Co dumbbells are formed.\cite{T9} While no ordering of the Fe and Co atoms was observed on the macroscopic level, the $^{69}$Ga NQR local probe revealed the primary formation of homo-atomic pairs Fe--Fe and Co--Co, although the Fe--Co dumbbells are also present in a significant amount.\cite{T9}

In this paper, we report the results of the systematic study of the evolution of the electronic structure and magnetic properties with Co substitution for Fe in the solid solution Fe$_{1-x}$Co$_x$Ga$_3$, including electrical resistivity ($x = 0.01$, 0.025, 0.075 and 0.5), magnetization ($x = 0.5$), \textit{ab initio} band structure calculations ($0\!\leq\!x\!\leq\!1$), and nuclear spin-lattice relaxation $1/T_1$ of the $^{69,71}$Ga nuclei. The electrical resistivity shows the anticipated metallic behavior at sufficiently high doping levels, namely, at and above $x\!=\!0.075$. Magnetization data for Fe$_{0.5}$Co$_{0.5}$Ga$_3$ are indicative of the localized magnetism with a relatively low effective moment. However, the nuclear spin-lattice relaxation $1/T_1$ of the $^{69,71}$Ga nuclei in the median compound Fe$_{0.5}$Co$_{0.5}$Ga$_3$ unexpectedly shows signatures of an itinerant antiferromagnetic behavior, with the $1/T_1$ strongly enhanced due to spin fluctuations. This scenario contrasts with the ferromagnetism observed in FeGa$_{3-y}$Ge$_y$ and implied by \textit{ab initio} band structure calculations for electron-doped FeGa$_3$. 

The nuclear spin-lattice relaxation of the binary CoGa$_{3}$ compound which is a band metal free from fluctuating spins demonstrates the metallic-like Korringa behavior with $1/T_1 \propto T$. In the parent semiconducting compound FeGa$_{3}$, the temperature dependence of the $1/T_1$ exhibits an unexpected huge maximum at low $T\approx 6$~K indicating the existence of in-gap states placed near the Fermi energy. These states seem to be responsible for the giant thermopower observed at the verge of magnetism in Fe based semimetals FeSb$_{2}$,\cite{T7,T29} Fe$_{1-x}$M$_{x}$Si alloys (M = Co, Ir, Os) \cite{T9,T30} and FeGa$_{3}$.\cite{T11} Surprisingly, we found that the in-gap states in FeGa$_{3}$ have a magnetic origin. Just recently in-gap states earned an increased attention because the question arises if these metallic and magnetic states are located at the surface and therefore have a topological origin. SmB$_{6}$ can be considered as a prototype of a topological correlated semimetal.\cite{T31}

\section{Experimental}

Single-phase powder samples of the Fe$_{1-x}$Co$_{x}$Ga$_{3}$ solid solution for various $x$ were prepared by mixing the elements (Fe: powder, Acros Organics 99\%; Co: powder, Alfa Aesar 99.8\%; Ga: bar, Aldrich 99.9999\%) with the Fe:Co:Ga molar ratio of $(1-x):x:15$ using Ga both as a reactant and flux medium. Mixtures of the elements were sealed in precarbonized quartz ampoules under vacuum (less than 10$^{-2}$ torr) and annealed in a programmable furnace at 900\,$^{\circ}$C for 55 hours to obtain a homogeneous melt. After this a furnace was slowly cooled to 400\,$^{\circ}$C in 125 hours and further to ambient temperature in 5 hours. Excess Ga was separated at 40\,$^{\circ}$C using an Eppendorf 5804R centrifuge, yielding needle-like silvery-gray crystals with a length up to several millimeters. The obtained crystals were purged from the remainder of Ga with diluted 0.5 M HCl and washed consecutively with distilled water and acetone. X-ray powder diffraction study (Bruker D8 Advance, Cu-K$\alpha$$_{1}$ radiation, Lynxeye detector) confirmed phase purity of all prepared samples.

Electrical resistivity measurements were performed using a standard four-probe method with the DC setup of a Quantum Design Physical Property Measurement System (PPMS) in the temperature range $4-300$ K in zero magnetic field. Electrical contacts (Cu 50 $\mu$m wire) were fixed on rectangular-shaped pellets with a typical size of $8\times 3\times 2$~mm$^3$ using silver-containing epoxy resin (Epotek H20E) hardened at 120\,$^{\circ}$C. Pellets were cold pressed from crystals ground in an agate mortar. Densities of the obtained pellets were estimated from their linear sizes and masses to be in the range of 89--92\% from the theoretical ones.

Magnetization measurements were performed on polycrystalline samples with a Quantum Design SQUID magnetometer MPMS-XL. DC magnetic susceptibility was measured in an applied field of 0.1~T in the $1.8-300$~K temperature range. The magnetization was measured at 1.8~K with the magnetic field varying from 0 to 7~T.

The $^{69,71}$Ga NMR/NQR measurements were performed in the temperature range $2-300$~K utilizing a home-built phase coherent pulsed NMR/NQR spectrometer. The $^{69,71}$Ga NQR spectra were measured using a frequency step point-by-point spin-echo technique. At each frequency point, the area under the spin-echo magnitude was integrated in the time domain and averaged by a number of accumulations, which depends on the sample and temperature. Nuclear spin-lattice relaxation rates were measured using the so-called ``saturation recovery'' method. Nuclear magnetization recovery curves $M(t)$ were obtained from the recovery of the spin-echo magnitude as a function of the time interval $\tau$ between the saturation pulse comb and the $\frac{\pi}{2}-\pi$ spin-echo sequence.

\section{Results and discussion}

\subsection{Electrical resistivity}

\begin{figure}
\center{\includegraphics[width=1\linewidth]{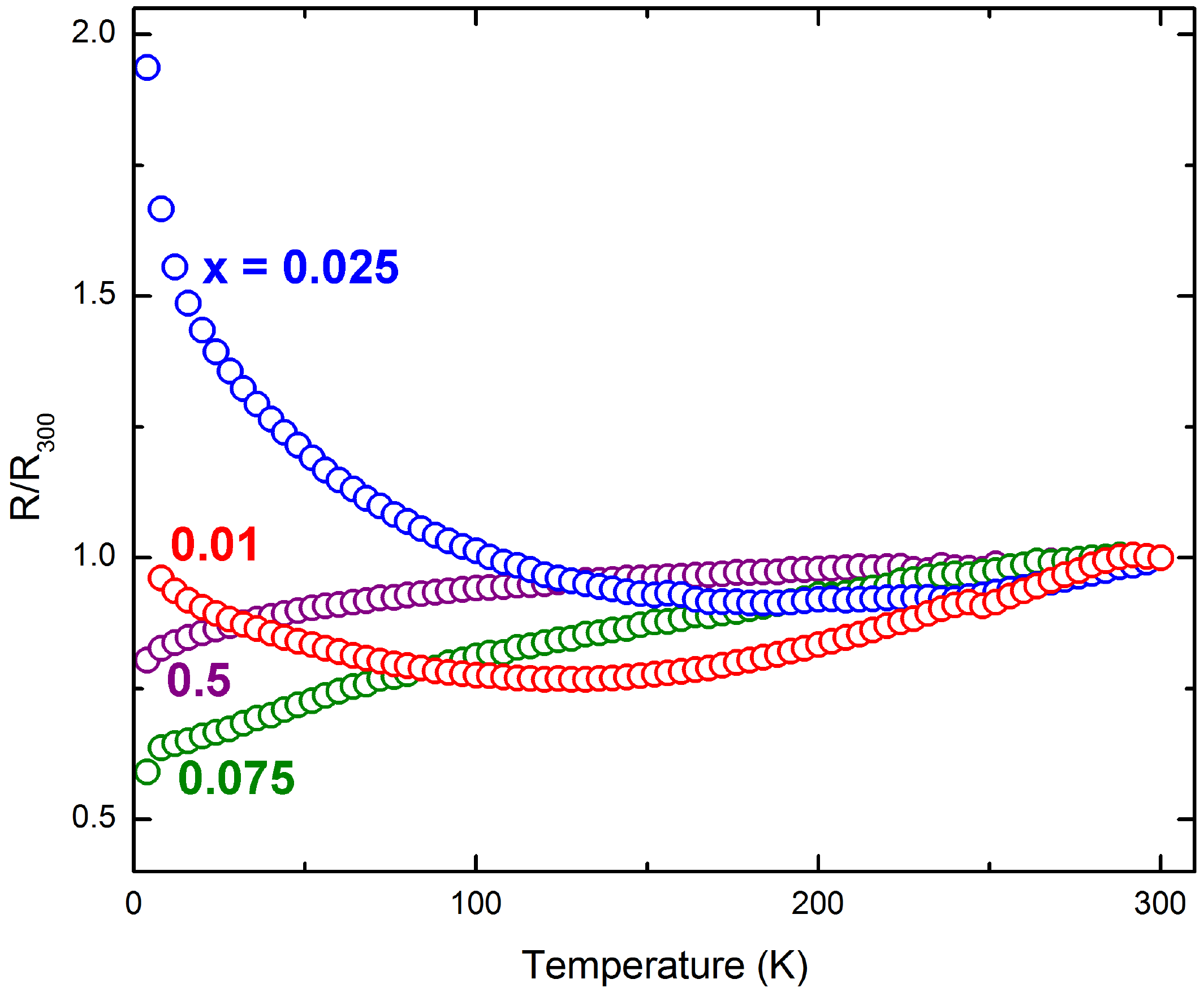}}
\caption{Normalized resistivity versus temperature data for the selected Fe$_{1-x}$Co$_{x}$Ga$_{3}$ specimens ($x=0.01$, 0.025, 0.075 and 0.5).}
\label{S2}
\end{figure}

Given the discrepancies in the earlier literature,\cite{T9, T15} we performed a systematic study of the electrical resistivity of the Fe$_{1-x}$Co$_{x}$Ga$_3$ solid solution starting from $x = 0.01$ (Figure~\ref{S2}). The samples with a small Co content ($x=0.01$ and 0.025) behave differently from those with $x\geq 0.075$. While at $x\geq 0.075$ the solid solution exhibits metallic behavior, a significant increase in the resistivity upon cooling below 150~K is observed for $x=0.01$ and 0.025. It is known from the literature\cite{T16} that the resistivity of pure FeGa$_3$ has different regimes in various temperature ranges. In comparison with FeGa$_3$, the behavior of the solid solution for $x=0.01$ and 0.025 can be described as follows: for $150<T<300$~K the saturation of thermally activated impurity donors is observed, while for $4<T<150$~K the impurity donors are frozen out. Such behavior suggests that the solid solution for $x=0.01$ and 0.025 should be regarded as a heavily-doped semiconductor. 

The transition from the semiconducting to metallic state occurs at $0.025\!<\!x\!<\!0.075$. This result agrees with our previous observation that at $x=0.125$ the solid solution is already metallic.\cite{T9} On the other hand, we do not confirm the results by Umeo~\textit{et al.},\cite{T15} who observed semiconductor-like behavior at $x=0.1$.

\subsection{\label{sec:level2}Magnetization}

The magnetic susceptibility for the Fe$_{0.5}$Co$_{0.5}$Ga$_{3}$ sample that was further studied with NQR (see below) was investigated. Similar to Ref.~\onlinecite{T15}, we observed an overall paramagnetic behavior. Above 100~K, the susceptibility follows the Curie-Weiss law:
\begin{equation}
\chi = \frac{C_{\sf CW}}{T - \theta_{\sf CW}} + \chi _0,
\label{F1}
\end{equation}
where $C_{\sf CW}$ is the Curie-Weiss constant, $\theta_{\sf CW}$ is the Curie-Weiss temperature, and $\chi_0$ stands for a temperature-independent contribution including core diamagnetism, van Vleck paramagnetism, and the contribution of itinerant electrons. We find $C_{\sf CW}=0.068$~emu\,K/mol, $\theta_{\sf CW}=-100$~K, and $\chi_{0}=1.7 \times 10^{-4}$~emu/mol. The effective magnetic moment is $\mu_{\sf eff}=0.74$~$\mu_B$ according to $C_{\sf CW} = N_{A}g^{2}\mu_{B}^{2}/k_{B}$ with $g=2$. These fitting results are rather stable with respect to the fitting range. We always find a sizable and negative (antiferromagnetic) $\theta_{\sf CW}$ as well as the effective moment of $0.6-0.8$~$\mu_B$, in good agreement with the earlier data by Umeo \textit{et al.},\cite{T15} who also report the negative $\theta_{\sf CW}$ for a range of Co concentrations. 

Despite the clear tendency of Fe$_{0.5}$Co$_{0.5}$Ga$_{3}$ toward an antiferromagnetic behavior, the susceptibility of Fe$_{0.5}$Co$_{0.5}$Ga$_3$ shows a positive deviation from the Curie-Weiss law below 100\,K. However, we did not observe any signatures of ferromagnetism in this compound. The magnetization isotherm measured at 1.8\,K is nearly linear, with only a weak bending that can be tentatively approximated by the Brillouin function augmented by an arbitrary linear term, which accounts for the increasing magnetization above the saturation of a spin-$\frac12$ paramagnet:
\begin{equation}
 M(H) = \alpha g\mu_B S\times \tanh\left( \frac{g \mu_{B} S}{k_{B}T}H\right) + \gamma H,
\label{F2}
\end{equation}
where we use $S=\frac12$, $g=2$, $T=1.8$~K, and $\alpha$ stands for the fraction of saturated paramagnetic moments. We find $\alpha=0.015$ (3\% out of 0.5 doped electrons/f.u.), indicating only a minor fraction of paramagnetic spins saturated in low fields. Indeed, the $\theta_{\sf CW}$ of $-100$~K implies sizable antiferromagnetic interactions that have to be overridden by the magnetic field before saturation is reached. 

Fe$_{0.5}$Co$_{0.5}$Ga$_{3}$, and the Fe$_{1-x}$Co$_{x}$Ga$_{3}$ solid solution in general, are remarkably different from their FeGa$_{3-y}$Ge$_y$ counterpart. The Ge-based compounds are ferromagnetic, with a positive $\theta_{\sf CW}$ and the sizable remnant magnetization observed at low temperatures. The doping with Co triggers an antiferromagnetic behavior, but does not induce any long-range magnetic order down to 2 K. In the following, we will substantiate these findings by analyzing the spin-lattice relaxation rate of the $^{69,71}$Ga nuclei.

\begin{figure}
\center{\includegraphics[width=1\linewidth]{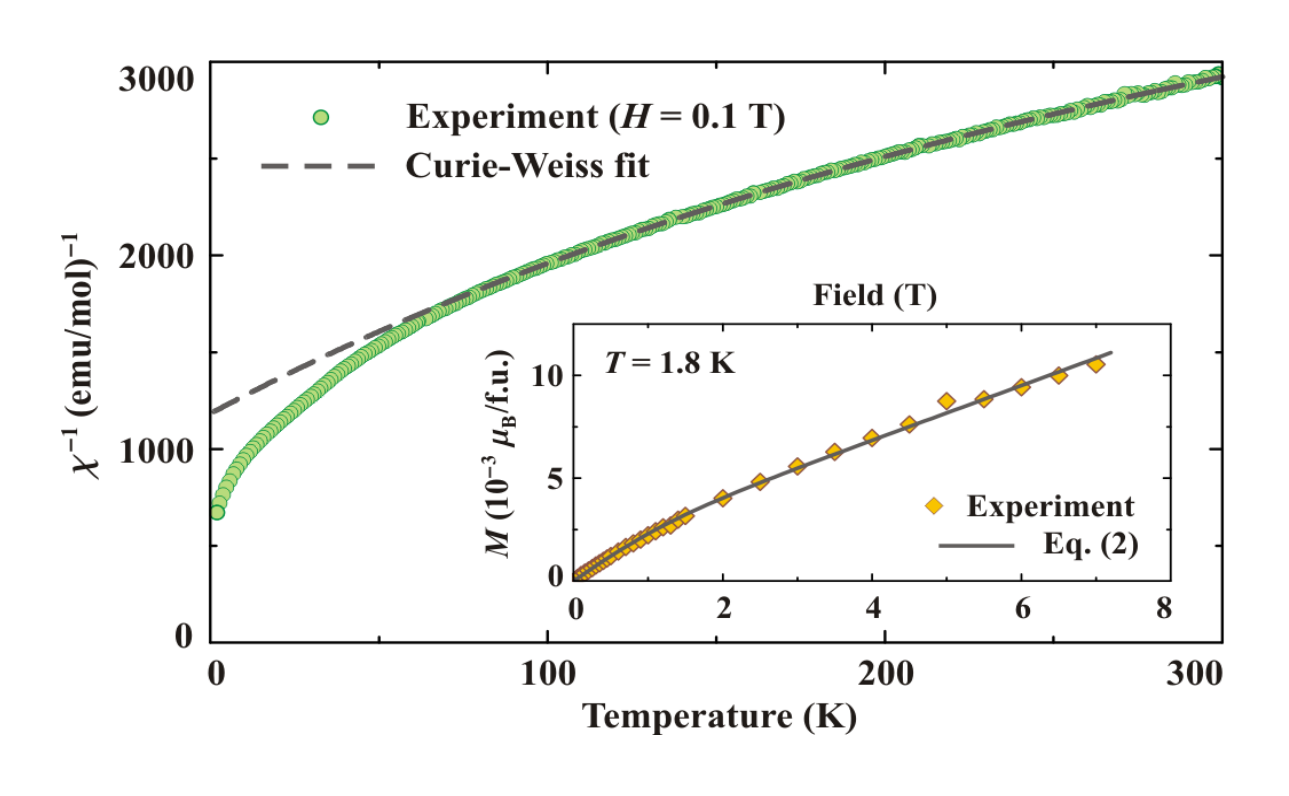}}
\caption{Inverse magnetic susceptibility of Fe$_{0.5}$Co$_{0.5}$Ga$_{3}$ measured in the applied field of 0.1~T. The dashed line is the fit with Eq.~\eqref{F1} above 100~K. The inset shows the field dependence of the magnetization, $M(H)$, at 1.8~K and its fit with Eq.~\eqref{F2}.}
\label{S3}
\end{figure}

\subsection{\textit{Ab initio} calculations}

The FPLO (full potential local orbitals) code was used for the electronic structure calculations.\cite{T18} FPLO performs density-functional (DFT) calculations within the local density approximation (LDA) for the exchange-correlation potential.\cite{T19} The integrations in the $k$ space were performed by an improved tetrahedron method\cite{T20} on a grid of $12\times 12\times 12$ $k$ points evenly spread in the first Brillouin zone. The calculation of the spin-polarized state of the Fe$_{1-x}$Co$_{x}$Ga$_3$ solid solution was done in the following steps. Firstly, an optimization of atomic coordinates was performed for different $x$ within the virtual crystal approximation (VCA) using the experimental values of lattice parameters.\cite{T9} And secondly, self-consistent spin-polarized calculations were performed for the relaxed structure. In addition to the VCA calculations, two types of ordering of the T atoms (T = Fe, Co) for $x=0.5$ were investigated. Using the results of the crystal structure determination for $x = 0.5$,\cite{T9} we constructed two models, the first one with homonuclear Fe--Fe and \mbox{Co--Co} dumbbells (space group $Cmmm$, $a=b=8.8298$~\r A, $c=6.4654$~\r A) and the second one with heteronuclear Fe--Co dumbbells (space group $Pmn2_1$, $a=6.4654$~\r A, $b=c=6.2436$~\r A). Spin-polarized states were calculated for these two models.

\begin{figure}
\center{\includegraphics[width=1\linewidth]{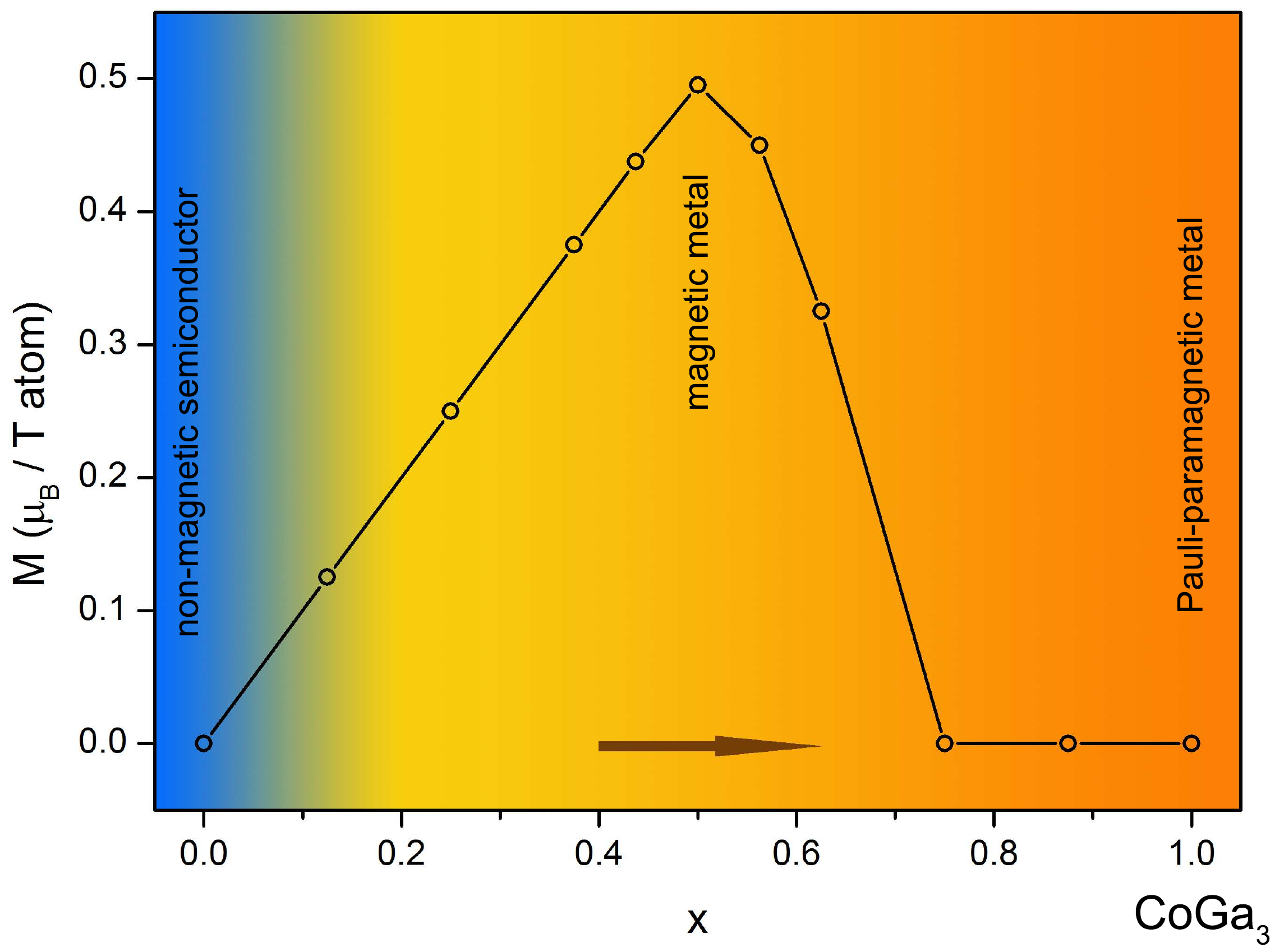}}
\caption{Total magnetic moment per T atom as a function of $x$ in Fe$_{1-x}$Co$_{x}$Ga$_{3}$ as obtained within the virtual crystal approximation.}
\label{S4}
\end{figure}

We considered both ferromagnetic and antiferromagnetic spin configurations. However, only the ferromagnetic configuration could be stabilized. A variety of antiferromagnetic spin patterns constructed within the unit cell of Fe$_{1-x}$Co$_{x}$Ga$_3$ and within doubled unit cells all converged to a non-magnetic solution lying higher in energy than the ferromagnetic solution. Therefore, DFT calculations on the LDA level put forward the ferromagnetic behavior of Fe$_{1-x}$Co$_{x}$Ga$_3$.
The variation of the total magnetic moment $M$ with $x$, as obtained within the virtual crystal approximation, is illustrated in Figure ~\ref{S4}. It shows that the maximal magnetic moment is achieved for $x = 0.5$. A similar result was obtained by Singh\cite{T21} for the same number of injected electrons upon the substitution of Ge for Ga in FeGa$_{3-y}$Ge$_y$. Indeed, on the VCA level the electronic structure of doped FeGa$_3$ evolves in a nearly rigid-band manner. Therefore, no difference between the Fe/Co and Ga/Ge doping should be expected.

\begin{figure}
\center{\includegraphics[width=1\linewidth]{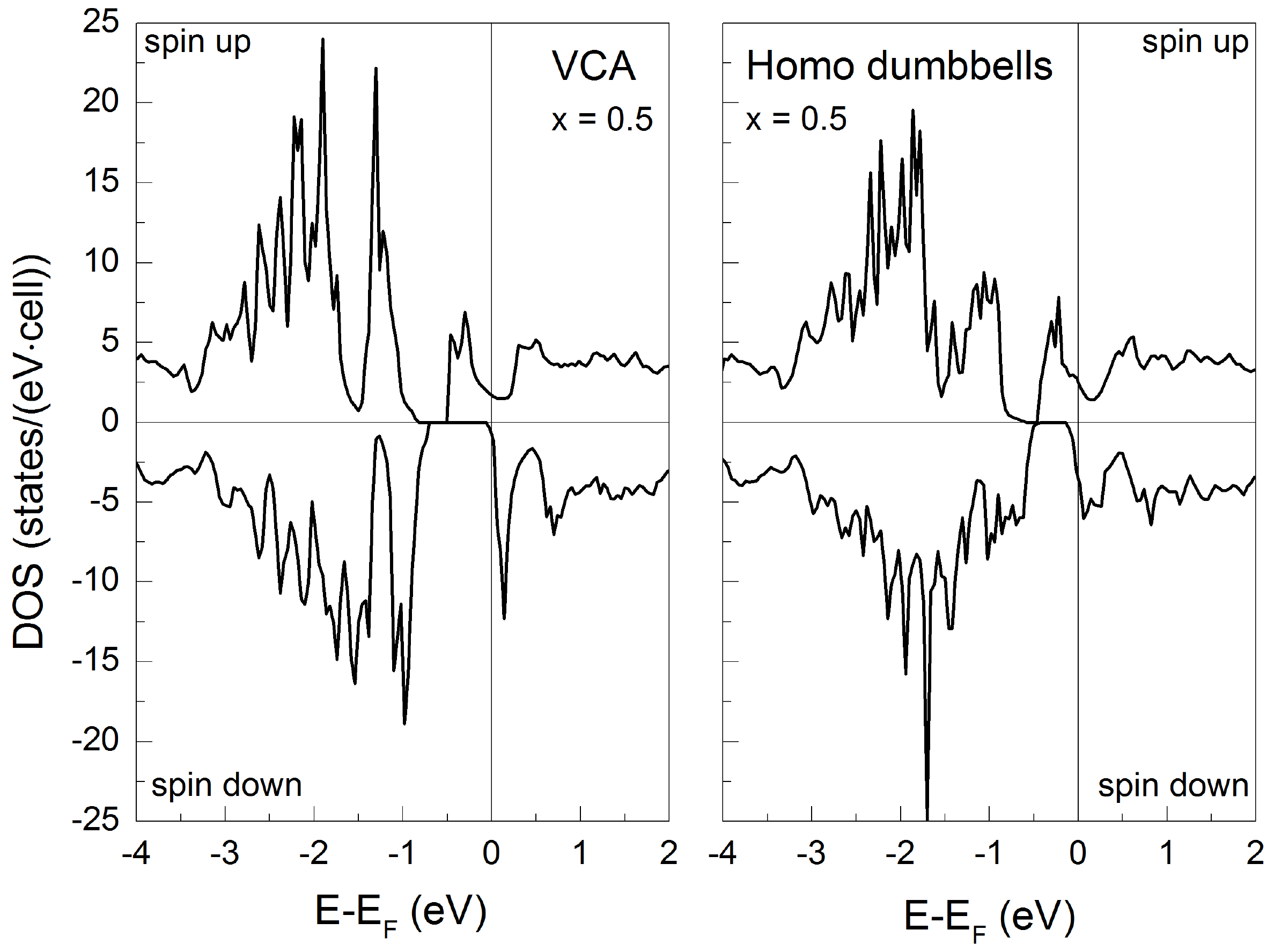}}
\caption{Density of states plots for $x=0.5$ calculated within the virtual crystal approximation (left) and for the ordered supercell with the homo-dumbbells Fe--Fe and Co--Co (right).}
\label{S5}
\end{figure}

The results of our calculations suggest that the half-metallic ferromagnetic state (Figure ~\ref{S5}, left) develops for $0<x\leq 0.5$. It turns to a metallic ferromagnetic state at $0.5<x<0.75$ and eventually becomes nonmagnetic at $0.75\leq x\leq 1$. This scenario is rather insensitive to the local order. The ordered supercells at $x=0.5$ also yield a ferromagnetic state with the local moment of 0.48~$\mu_B$ for the heteronuclear (Fe--Co) dumbbells and 0.41~$\mu_B$ for the homonuclear (Fe--Fe, Co--Co) dumbbells (Figure ~\ref{S5}, right).

The ferromagnetic ground state is indeed observed in FeGa$_{3-y}$Ge$_{y}$. The experimental ordered moment of 0.27~$\mu_B$/f.u. at $y=0.41$ is somewhat lower than the calculated one ($\mu_{\sf LDA} \sim 0.4$~$\mu_{B}$/f.u. at $y=0.4$), as typical for itinerant magnets, where spin fluctuations, which are missing in LDA, lead to a substantial reduction in the ordered moment. In the case of Fe$_{1-x}$Co$_{x}$Ga$_3$, the lack of magnetic ordering prevents us from a direct comparison of $\mu_{\sf LDA}$ to the experiment. Moreover, we find a qualitative difference between the ferromagnetic behavior, as predicted by LDA, and the antiferromagnetic behavior, which is evidenced by the negative $\theta_{CW}$ and additionally supported by the spin-lattice relaxation presented below. 

\subsection{NMR in Fe$_{1-x}$Co$_{x}$Ga$_{3}$}

\begin{figure}
\center{\includegraphics[width=1\linewidth]{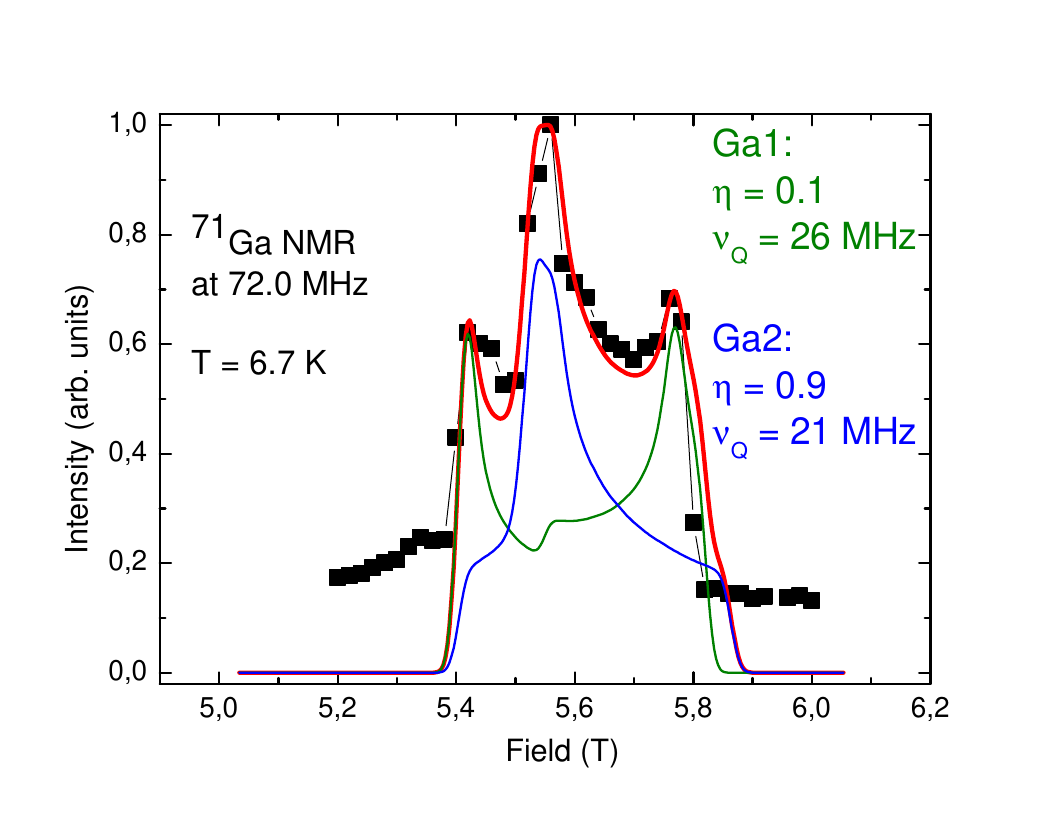}}
\caption{(Color online) FeGa$_3$: field-sweep NMR spectrum of the central ($-\frac12\leftrightarrow +\frac12$) transition of the $^{71}$Ga nuclei measured at $T=6.7$~K at 72.0~MHz. Thin solid lines (blue and green) are the numerical simulation for the Ga1 site ($\eta=0.1$, $\nu_Q=26.0$~MHz) and Ga2 site ($\eta=0.9$, $\nu_Q=21.8$~MHz). The thick red line is the resulting simulation spectrum.}
\label{S6}
\end{figure}

There are three NMR active and naturally abundant isotopes in Fe$_{1-x}$Co$_{x}$Ga$_{3}$ compounds suitable for NMR investigation: $^{69}$Ga, $^{71}$Ga, and $^{59}$Co. Our previous $^{69,71}$Ga NQR investigations\cite{T9} revealed high values of the quadrupole frequencies for both Ga positions in the Fe$_{1-x}$Co$_{x}$Ga$_3$ solid solution series. Two inequivalent positions of Ga lead to very broad $^{69,71}$Ga NMR spectra with overlapping singularities of the powder pattern. Thus, for the parent semiconductor compound FeGa$_3$ exhibiting the strongest quadrupole splitting, we have measured only the central transition line of the field-sweep NMR spectrum of the $^{71}$Ga isotope, which has a lower quadrupole frequency (Fig.~\ref{S6}). This spectrum is a textbook example of the NMR central transition powder pattern in the presence of quadrupole interactions in the second order of perturbation theory. It consists of two contributions: the characteristic double horn line with a step in the middle originating from the Ga1 position with a low asymmetry parameter $\eta=0.1$ (green line) and the single peak with typical shoulders at the edges, which stems from the Ga2 site with the high asymmetry parameter $\eta=0.9$ (blue line). One can see that the observed $^{71}$Ga NMR spectrum coincides very well with our NQR results reported in Ref.~\onlinecite{T9}.

\begin{figure}
\center{\includegraphics[width=1\linewidth]{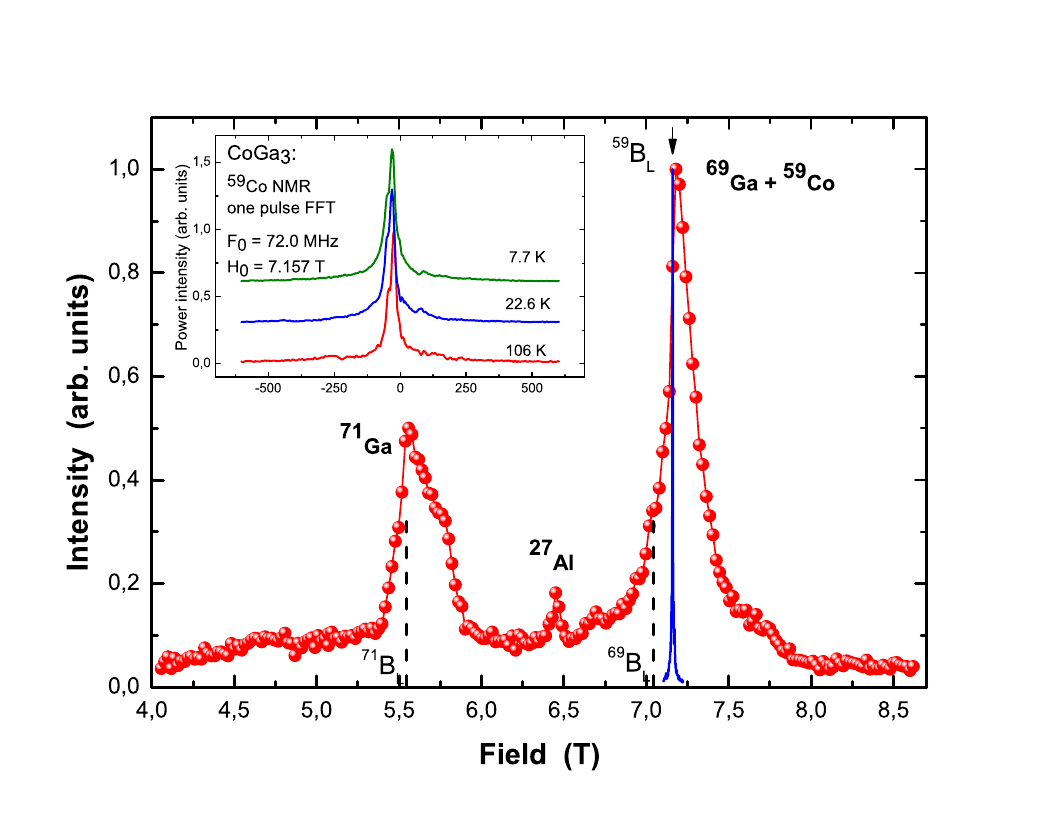}}
\caption{(Color online) Field-sweep NMR spectrum of Fe$_{0.75}$Co$_{0.25}$Ga$_{3}$ measured at 72.0~MHz at 19~K (red spheres). Vertical dashed lines indicate the position of the Larmor fields for $^{71}$Ga and $^{69}$Ga. The arrow shows the position of the Larmor field for $^{59}$Co. Blue solid line shows $^{59}$Co one pulse FT NMR spectrum measured at 22.6~K at $H=7.157$~T in CoGa$_3$ recalculated to the field domain. Also the position of the ghost $^{27}$Al line is shown. Inset: $^{59}$Co one pulse NMR FFT spectra measured at a fixed magnetic field of 7.157~T and a reference frequency of 72.0~MHz.}
\label{S7}
\end{figure}

With the Co substitution for Fe in Fe$_{1-x}$Co$_{x}$Ga$_{3}$, the $^{71}$Ga NMR central transition line broadens and loses its sharp singularities, reflecting the increasing inhomogeneity of the electric field gradient (EFG) distribution, in perfect agreement with the broadening of Ga NQR lines. This is demonstrated in Fig.~\ref{S7}, where the field-sweep spectrum of Fe$_{0.75}$Co$_{0.25}$Ga$_3$ is shown. Interestingly, a strong line appears in the field range of $7.0-7.5$~T. This line can be assigned to a $^{59}$Co NMR signal with some broad background from the $^{69}$Ga isotope.

For the opposite side binary compound CoGa$_3$, the NMR spectrum is completely dominated by an extremely narrow $^{59}$Co line observed very close to the Larmor field position of the $^{59}$Co nuclei at 72.0~MHz (Fig.~\ref{S7}). To measure such a narrow line properly, we had to reduce our spin echo pulse sequence to one pulse and use the Fast Fourier Transform (FFT) technique at a fixed magnetic field of 7.157\,T. The FFT $^{59}$Co spectra measured for CoGa$_3$ at various temperatures are presented in the inset of Fig.~\ref{S7}. The striking narrowness of the $^{59}$Co NMR line shows that the Co atom is in a completely nonmagnetic state, donating all 9 valence electrons to the conduction band, which results in good metallic properties of CoGa$_3$.\cite{T14}

The central transition line of the $^{59}$Co NMR spectra in CoGa$_3$ exhibit relatively low 2-nd order quadrupole splitting of about 50 kHz. The reason for this effect might be an effective dynamic screening of the lattice contribution to the electric field gradient (EFG) by conduction electrons, while the on-site contribution to the EFG is almost zero due to an absence of electrons localized on the valence $3d$ and $4s$ shells.

\subsection{Nuclear spin-lattice relaxation (NSLR)}

\begin{figure}
\center{\includegraphics[width=1\linewidth]{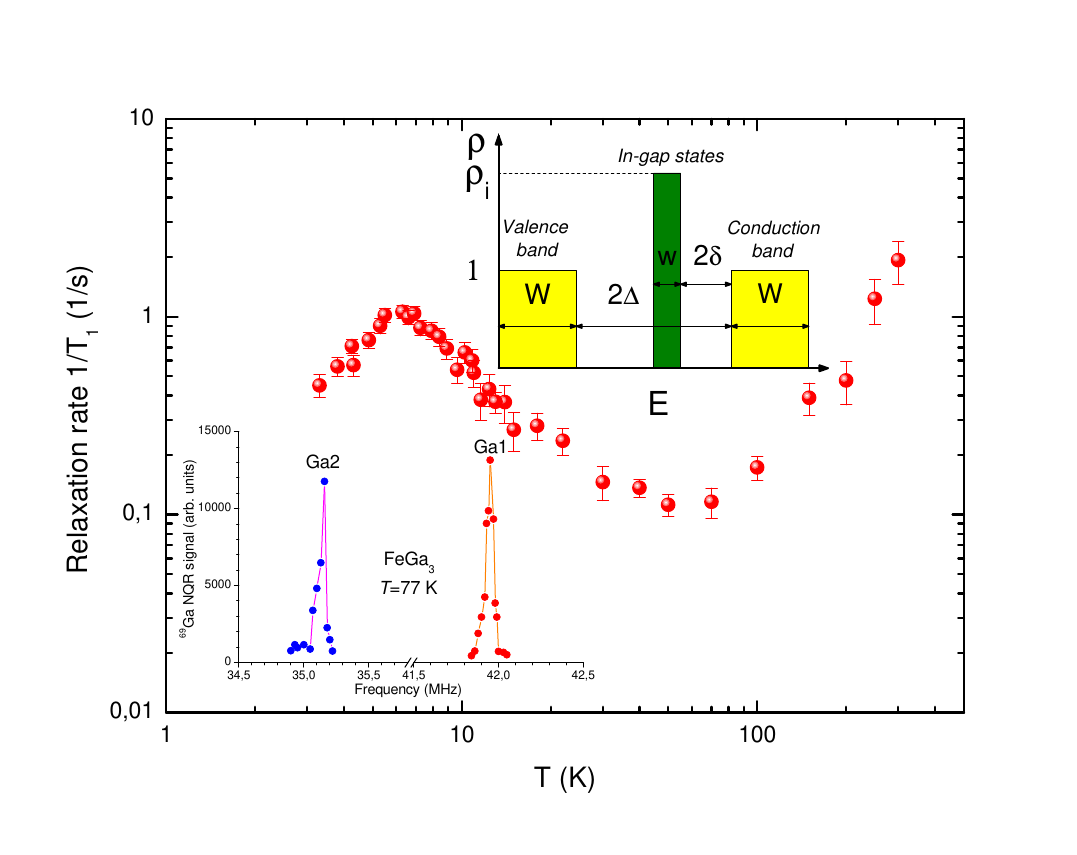}}
\caption{(Color online) Temperature dependence of the $^{69}$Ga NSLR at the Ga1 site in FeGa$_3$. Bottom inset: the $^{69}$Ga NQR spectrum in FeGa$_3$ measured at 77~K. Top Inset: modified ``narrow band -- small energy gap'' model (see text).}
\label{S8}
\end{figure}

\begin{figure}
\center{\includegraphics[width=1\linewidth]{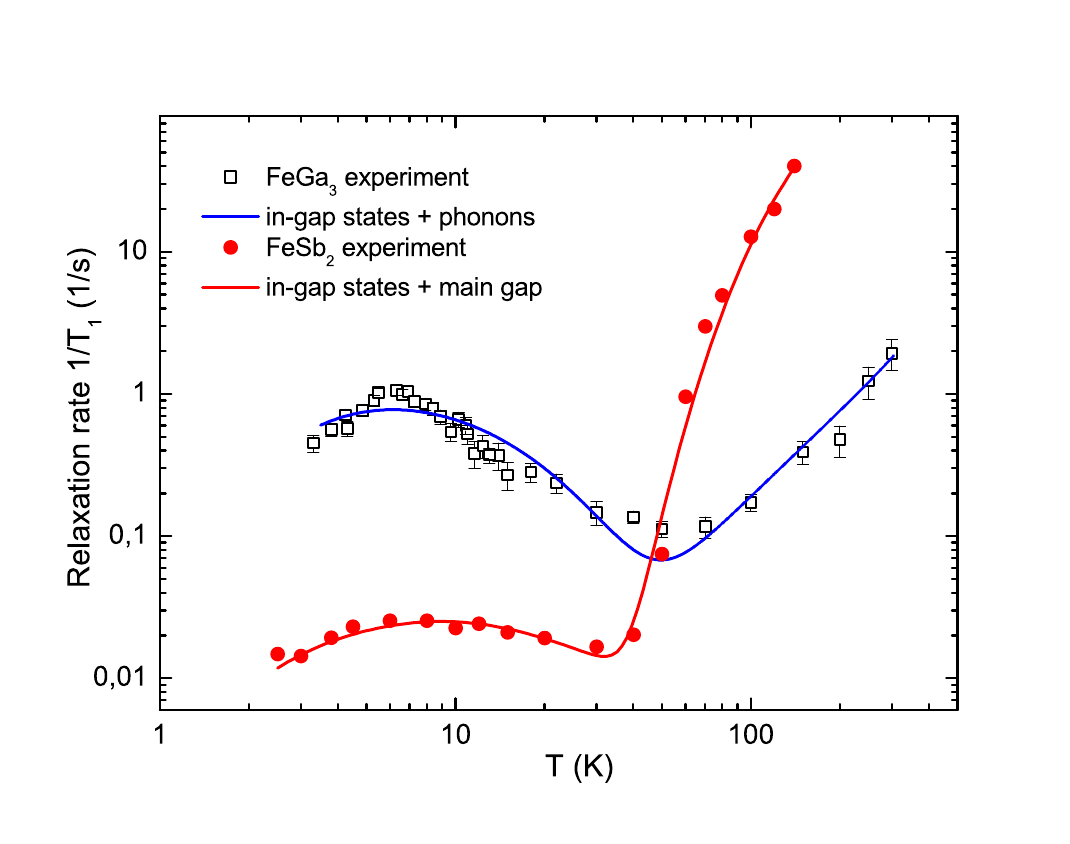}}
\caption{(Color online) Comparison of the $^{69}$Ga spin-lattice relaxation rate in FeGa$_3$ (open squares) and the $^{123}$Sb relaxation rate in FeSb$_2$ (closed circles, adopted from Ref.~\onlinecite{T5}). Solid lines are best fits using the model described in the text.}
\label{S9}
\end{figure}

The nuclear spin-lattice relaxation rate (NSLR) $1/T_1$ was measured using the $^{69}$Ga nuclei for the Ga1 NQR line (see the bottom inset in Fig.~\ref{S8}) in a wide temperature range of $3-300$~K for three Fe$_{1-x}$Co$_{x}$Ga$_{3}$ samples with the Co concentrations $x=0$, 0.5, and 1.0. For the edge binary compounds, the magnetization recovery curves were single exponential, while for Fe$_{0.5}$Co$_{0.5}$Ga$_{3}$ the stretched exponent function provides better fitting results. The temperature dependence of $^{69}(1/T_1)$ in FeGa$_3$ is presented in Fig.~\ref{S8}. Surprisingly, it demonstrates a huge unexpected maximum at $T\approx 6$~K with almost a one order of magnitude difference between the $1/T_{1}$ values at 6~K and at 50~K. This maximum unambiguously proves the existence of the in-gap states just below the conduction band. A very similar behavior was observed earlier for the $^{123}$Sb NSLR in FeSb$_2$.\cite{T5} The comparison of these NSLR data for FeGa$_{3}$ and FeSb$_{2}$ is presented in Fig.~\ref{S9}. 

For a quantitative description of our $^{69}(1/T_1)$ data in FeGa$_3$ we used the modified ``narrow band--small energy gap'' model.\cite{T22} It contains two rectangular bands of width $W$ separated by a main gap of $2\Delta$ with a narrow peak of in-gap states of width $w$ with a density of states $\rho_i (\varepsilon)\!\sim\!\rho_{0}\exp(-T/T_{0}$) separated by the small gap of $2\delta$ from the bottom of the conduction band (see Fig.~\ref{S8}, Top Inset). There are two main temperature regimes in this model. In the low-temperature (LT) regime, the NSLR mechanism is caused by the activation of the localized in-gap electrons over a small gap $2\delta$ into the empty conduction band. This leads to the gradual increase of Ga 1/$T_1(T)$ with increasing temperature from the lowest $T\sim 3$\,K to the temperature of the NSLR maximum around 6\,K.  The narrow in-gap peak $\rho_{i}(\varepsilon)$ disappears at higher temperatures due to its broadening and overlap with the conduction band, resulting in the decreasing NSLR and in the appearance of a clearly pronounced minimum on $1/T_{1}(T)$ at 50--60~K. Then, in the high-temperature (HT) regime, $1/T_{1}$ starts increasing again due to the electron activation across the main energy gap $2\Delta$.

In general, the NSLR can be expressed as:
\begin{equation}
\frac{1}{T_1} = \frac{\hbar k_B T}{\omega} \sum \limits_{q} A^2 F(q) \chi_{q,w}'',
\label{F3}
\end{equation}
where $\chi_{q,w}''\approx \frac{\pi}{k_B T} \int d\varepsilon f(\varepsilon) [1 - f(\varepsilon)] \rho^2 (\varepsilon)$  is the imaginary part of the dynamical susceptibility; 
$f(\varepsilon) = \left(\exp(\varepsilon/T)+1\right)^{-1}$ is the Fermi distribution function, $A$ is the hyperfine interaction constant, $F(q)$ is the form-factor depending on the geometry. Assuming a weak $q$- and $\omega$- dependence of $A$ and $F(q)$ and using the equation $-T\delta f / \delta \varepsilon = f(\varepsilon)(1-f(\varepsilon)$, one obtains after integration of Eq.~\eqref{F3} over the rectangular bands:
\begin{equation}
\frac{1}{T_1} \propto T\left[\rho_d ^2\left(f(\Delta) - f(\Delta + W)\right) + \rho _i ^2 \left( f(\delta)-f(\delta+\omega)\right)\right]
\label{F4}
\end{equation}

Here, the first term is responsible for the activation-like relaxation behavior in the HT regime, while the second one causes the pronounced maximum in the LT regime. 
To complete the relaxation scenario, one should add to Eq.~\eqref{F4} the phonon-induced quadrupole relaxation involving two-phonon (Raman) scattering, which is active already at moderate temperatures. Indeed, the EFG at the Ga site is strongly affected by thermal fluctuations. The interaction of the fluctuating EFG with the quadrupole moments of the Ga nuclei causes the quadrupole relaxation, for which the relaxation rate increases with temperature as $T^2$ (see Refs.~\onlinecite{T23,T24}):
\begin{equation}
\frac{1}{T_1}\propto \frac{9k_B ^2}{8 \pi ^3 \hbar ^3} \frac{e^2 \gamma Q}{R^3}\, T^2,
\label{F5}
\end{equation}
where $\gamma$ is the gyromagnetic ratio, $Q$ the quadrupole moment, and $R$ the interatomic distance. Adding Eq.~\eqref{F5} in a more general power form 1/$T_1$ = $AT^n$ to the expression~\eqref{F4}, one arrives for FeGa$_3$ at: 

\begin{eqnarray}
\frac{1}{T_1} \propto &T\left[\rho _d^2\left( f(\Delta)-f(\Delta+W)\right)\right.+ \notag\\
&\,\, +\rho_i^2\left.\left( f(\delta)-f(\delta+\omega)\right)\right] + AT^n
\label{F6}
\end{eqnarray}

Using Eq.~\eqref{F6}, we succeeded to fit the experimental $^{69}$Ga $1/T_1$ data in the entire investigated temperature range $3-300$~K. The best fit of the experimental $1/T_1$ data to Eq.~\eqref{F6} (blue solid line in Fig.~\ref{S9}) gives the power factor $n\approx 2$ confirming the $T^2$ behavior characteristic for the phonon relaxation mechanism driven by the two-phonon (Raman) scattering.

For comparison, we used Eq.~\eqref{F6} to fit the experimental $^{123}$Sb $1/T_1(T)$ data for FeSb$_2$ between $T=2.5-150$~K adapted from Ref.~\onlinecite{T5}. The best fitting curve for FeSb$_2$ is shown in Figure~\ref{S9} by the red solid line. The resulting fitting parameters for FeSb$_2$ and FeGa$_3$ are presented in Table~\ref{B1}. As seen from these values, the in-gap states in both compounds indeed form a very narrow layer with a width of $w=1$\,K (26\,K) separated from the bottom of the conduction band only by $2\delta=13$\,K (8\,K) for FeGa$_3$ (FeSb$_2$).

\begin{table*}
\caption{\label{B1}Best fit parameters according to the modified ``Narrow band -- small energy gap'' model\cite{T22} (see text).}
\begin{ruledtabular}
\center{\begin{tabular}{p{0.7\linewidth}p{0.15\linewidth}p{0.15\linewidth}}
Parameter & FeGa$_3$ & FeSb$_2$\\
In-gap layer separation from the conduction band, 2$\delta$ & 13 K  & 8 K \\
Main gap, $2\Delta$ & 550 K  & 800 K\\
In-gap layer width, $w$ & 1 K & 26 K\\
Main band width, $W$ & 75 K & 1100 K\\
Zero $T$ height of the in-gap layer, normalized to the main rectangle band height, $\rho_0$ & 1.73 K$^{-1}$ & 0.06 K$^{-1}$\\
Relative zero $T$ capacity of the in-gap layer, $n_0$ = ($\rho_0 w)/(\rho_W W$) & $2.3\times 10^{-2}$ & $1.4\times 10^{-3}$\\
\end{tabular}}
\end{ruledtabular}
\end{table*}

It is worth comparing the relative capacity of the in-gap state level in FeGa$_3$ and FeSb$_2$ estimated as $n_0$ = ($\rho_{0}w$)/($\rho_{W}W$), where $\rho_W$ = 1 is the normalized height of the main rectangular bands. As seen from Table~\ref{B1}, this value is about 16 times higher in FeGa$_3$ than in FeSb$_2$. This explains the much stronger and more pronounced $1/T_1(T)$ maximum at low temperatures observed in FeGa$_3$ in comparison to that in FeSb$_2$ (Fig.~\ref{S9}).
The obtained main gap $E_g=2\Delta$ value in FeGa$_3$ is in perfect agreement with our \textit{ab initio} calculations $2\Delta_{ab initio} \approx$ 0.4 eV ($\approx$ 4700 K).\cite{T9} From this result, one can expect different scenarios for $1/T_{1}(T)$ in the HT regime for FeGa$_{3}$ and FeSb$_{2}$. Actually, the energy gap value of $2\Delta$ for FeGa$_3$ is nearly 7 times larger than that for FeSb$_2$: 5500\,K vs. 800\,K, respectively. This means that in the investigated HT regime 40--300\,K the thermal activation of electrons across the main gap is rather inefficient for FeGa$_3$ yet.

Summarizing the above consideration, the spin-lattice relaxation in FeGa$_3$ and FeSb$_2$ can be decomposed into three parts:
\begin{equation}
\left(\frac{1}{T_1}\right)_\Sigma = \left(\frac{1}{T_1}\right)_{\sf in-gap} + \left(\frac{1}{T_1}\right)_{\sf act} + \left(\frac{1}{T_1}\right)_{\sf ph},
\label{F7}
\end{equation}
where $\left(T_1^{-1}\right)_{\sf in-gap}$ is the magnetic relaxation caused by activation from the in-gap states into conduction band; $\left(T_1^{-1}\right)_{\sf act}$  is the magnetic relaxation due to activation of $3d$ electrons over the main energy gap $E_g=2\Delta$, and $\left(T_1^{-1}\right)_{\sf ph}$ is the quadrupole relaxation caused by phonons. 

The interplay between these components determines the observed temperature behavior of the spin-lattice relaxation in FeGa$_3$ and FeSb$_2$. At low temperatures ($T<40$\,K), the first term of Eq.~\eqref{F7} gives the main contribution to $1/T_1$ in both FeGa$_3$ and FeSb$_2$ providing a pronounced maximum of $1/T_1(T)$ at $T_{\max}\approx 6$\,K and 10\,K, respectively. With increasing temperature, other terms in Eq.~\eqref{F7} start to dominate causing the fast increase in $1/T_1$. In FeSb$_2$, the energy gap $E_g$ is relatively small and the main relaxation channel is an activation with an exponential growth of $1/T_1$. In contrast, FeGa$_3$ exhibits a much larger gap value making the activation process ineffective, which results in a dominance of the phonon relaxation in the temperature range 40--300\,K. 

\begin{figure}
\center{\includegraphics[width=1\linewidth]{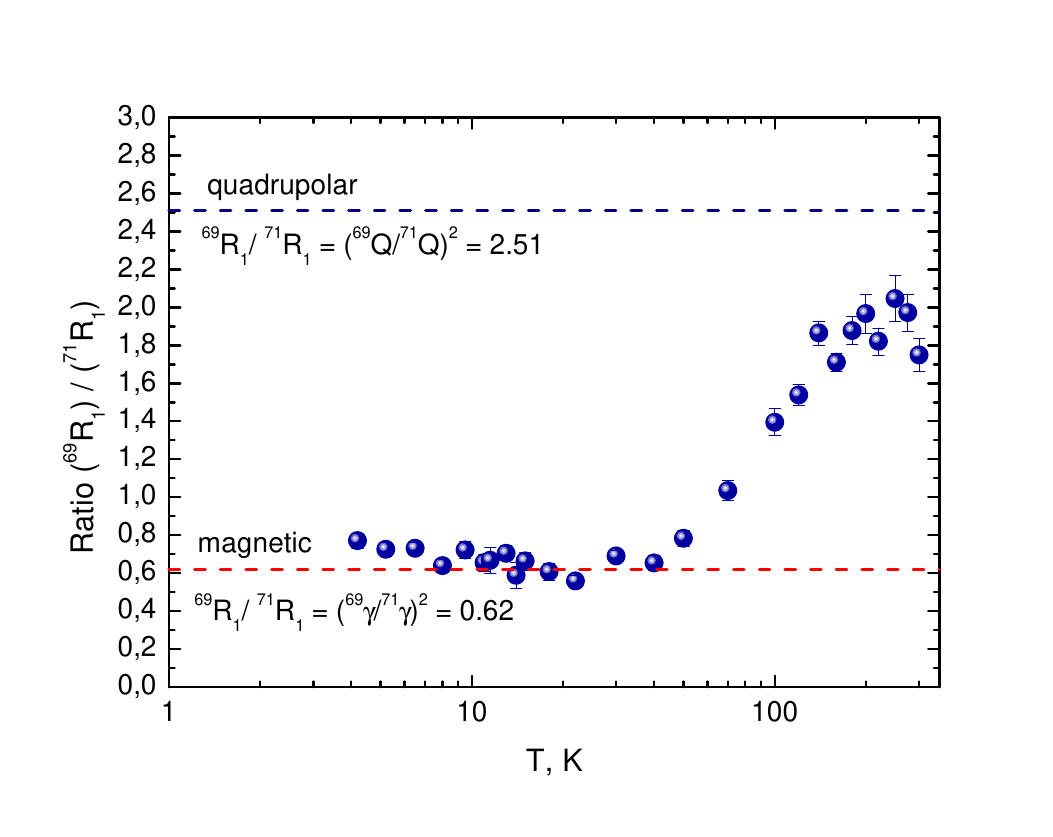}}
\caption{(Color online) Ratio of the spin-lattice relaxation rates for $^{69}$Ga and $^{71}$Ga as a function of temperature in FeGa$_3$. Dashed lines indicate the values for pure quadrupolar and pure magnetic relaxation mechanism.}
\label{S10}
\end{figure}

This relaxation scenario in FeGa$_3$ is independently confirmed by the isotope effect analysis. In the case of a pure magnetic relaxation mechanism, the ratio of the relaxation rates $^{69}(1/T_1)/^{71}(1/T_1)$ of the $^{69}$Ga and $^{71}$Ga isotopes should be equal to $(^{69}\gamma /^{71} \gamma)^2=0.62$, where $\gamma$ is the gyromagnetic ratio of the corresponding nuclei, while in the opposite case of a pure quadrupole relaxation mechanism $^{69}(1/T_1)/^{71}(1/T_1)=(^{69}Q/^{71}Q)^2=2.51$, where $Q$ is the quadrupole moment of the nuclei. The experimental $^{69}(1/T_1)/^{71}(1/T_1)$ data as a function of temperature are shown in Fig.~\ref{S10}. As clearly seen from this figure, in the low-temperature range from 2\,K to 40\,K the relaxation mechanism is exclusively magnetic. With further increase in temperature, the quadrupolar contribution to the total relaxation rate increases rapidly. Finally, above 100\,K one arrives at the mixed case of both magnetic and quadrupole channels of nuclear relaxation with the pronounced domination of the latter. This result is in good agreement with the analysis described above.

\begin{figure}
\center{\includegraphics[width=1\linewidth]{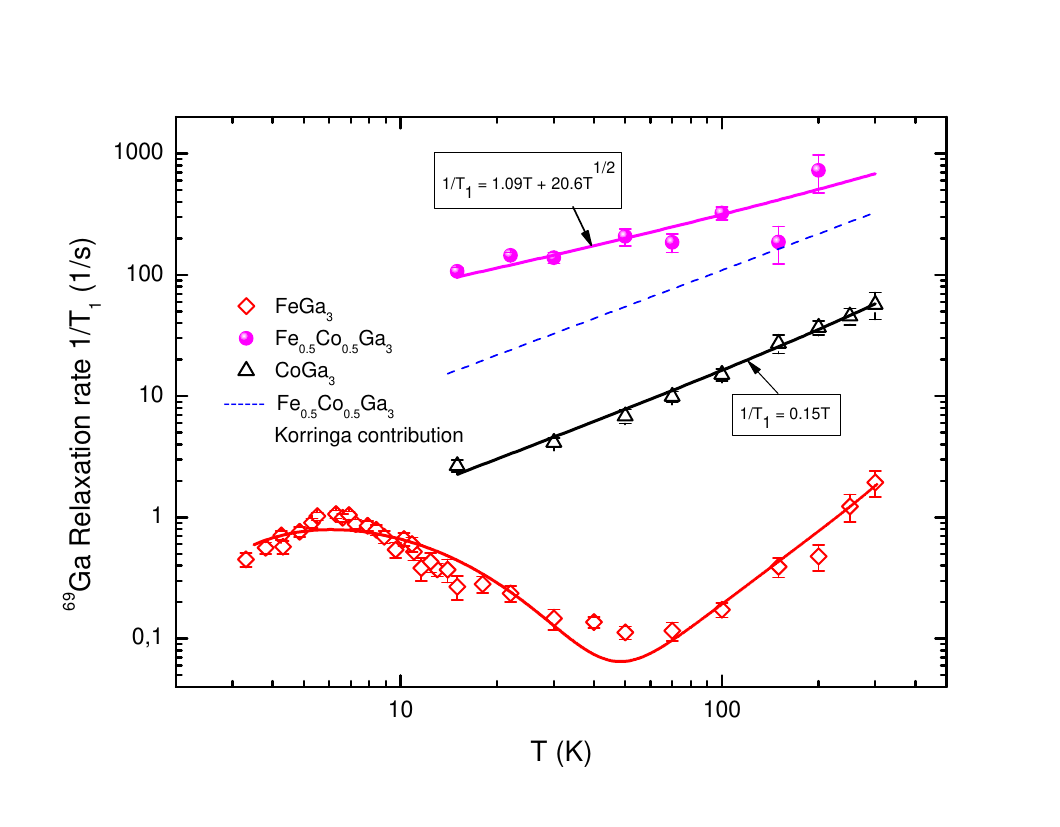}}
\caption{(Color online) Temperature dependence of the spin-lattice relaxation rate $1/T_1$ for FeGa$_3$, CoGa$_3$, and Fe$_{0.5}$Co$_{0.5}$Ga$_3$. Solid lines are best fits using the models described in the text.}
\label{S11}
\end{figure}

Due to its good metallic properties, the opposite edge binary compound CoGa$_3$ exhibits a Korringa-like linear $^{69}$Ga spin-lattice relaxation caused by the contact interaction between the conduction electrons and $^{69}$Ga nuclei:
\begin{equation}
\left(\frac{1}{T_1}\right)_K = \pi \hbar ^3 \gamma _e ^2 \gamma _n ^2 A _{\sf hf} ^2 N^2 (E_F) k_B T = a_KT,
\label{F8}
\end{equation}  
where $\gamma_e$ and $\gamma_n$ are the gyromagnetic ratios for electron and nucleus, $A_{\sf hf}$ is the contact hyperfine coupling between conduction electrons and nuclei, $N(E_F)$ is the density of states at the Fermi level, and $a_K$ is the Korringa coefficient. As seen from Figure~\ref{S11}, the experimental $^{69}(1/T_1(T))$ data for CoGa$_3$ can be perfectly approximated by Eq.~\eqref{F8} in the entire investigated temperature range of 12--300\,K with the best fit value of $a_K=0.15$.

A very fast relaxation rate is shown by the Fe$_{0.5}$Co$_{0.5}$Ga$_3$ compound: more than one order of magnitude faster than in CoGa$_3$ (Fig.~\ref{S11}). At first glance, it is quite natural since according to our band structure calculations\cite{T9} the ratio of the squared densities of states at the Fermi level for Fe$_{0.5}$Co$_{0.5}$Ga$_3$ and CoGa$_3$ $N_{E_F}$(Fe$_{0.5}$Co$_{0.5}$Ga$_3$)/$N_{E_F}$(CoGa$_3$) = 7.26. Therefore, the linear Korringa coefficient for Fe$_{0.5}$Co$_{0.5}$Ga$_3$ can be estimated as:
\begin{equation}
a_K(\text{Fe}_{0.5}\text{Co}_{0.5}\text{Ga}_3 ) = 7.26\times a_K(\text{CoGa}_3) = 1.09.
\label{F9}
\end{equation}  
The linear Korringa function $1/T_1=1.09\times T$ is plotted in Fig.~\ref{S11} by the dashed line. It is indeed above the relaxation data for CoGa$_3$ but still far below the experimental 1/$T_1$ data for the Fe$_{0.5}$Co$_{0.5}$Ga$_3$ compound. Moreover, it has a wrong decline in the double logarithmic scale of Fig.~\ref{S11}. This result unambiguously shows that the spin-lattice relaxation in Fe$_{0.5}$Co$_{0.5}$Ga$_3$ is dominated by a mechanism other than the Korringa mechanism. The model which describes our relaxation data in Fe$_{0.5}$Co$_{0.5}$Ga$_3$ is the Moriya's spin-fluctuation theory.\cite{T25} According to this theory for weak (low $T_N$) or nearly antiferromagnetic (AF) metals the nuclear spin-lattice relaxation is given by the equation:
\begin{equation}
\frac{1}{T_1} = \frac{\alpha T}{(T-T_N ) ^{1/2} }.
\label{F10}
\end{equation}                                                                                                                    
For $T\!\gg\!T_N\!\approx\!0$, Eq.~\eqref{F10} is reduced to a square root temperature dependence $1/T_1\propto T^{1/2}$, which is a unique feature of weakly and nearly AF metals.\cite{T25} As shown in Fig.~\ref{S11}, our experimental $1/T_1$ data for Fe$_{0.5}$Co$_{0.5}$Ga$_3$ can be successfully fitted by a combination of the linear Korringa term and spin-fluctuation contribution with dominance of the latter:
\begin{equation}
\frac{1}{T_1} = 1.09 \times T + 20.6 \times T^{1/2}
\label{F11}
\end{equation}   
This result is very similar to that observed by $^{55}$Mn NMR in $\beta$-Mn, which is an AF metal subject to a strong magnetic frustration ($T_N\approx 0$).\cite{T26} The $^{55}$Mn relaxation in $\beta$-Mn is mainly due to spin fluctuations, with a minor contribution from contact Korringa interactions: $1/T_1$ = 1.7$\times T$ + 35.3$\times T^{1/2}$. Pure spin-fluctuation scenario of nuclear spin-lattice relaxation with $1/T_1\propto T^{1/2}$ in the wide temperature range of 2--300\,K was observed by $^{139}$La NMR in filled skutterudites La$_{0.9}$Fe$_4$Sb$_{12}$.\cite{T27}

\section{Summary and conclusions}

Our experimental and computational study puts forward the complex magnetic behavior of the Fe$_{1-x}$Co$_x$Ga$_3$ solid solution. According to the NSLR data, the median compound Fe$_{0.5}$Co$_{0.5}$Ga$_3$ reveals strong spin fluctuations that are characteristic of a weakly antiferromagnetic metal. This observation is consistent with the negative (antiferromagnetic) $\theta_{\sf CW}$ obtained in the Curie-Weiss fit of the magnetic susceptibility. However, the $1/T$ (Curie-Weiss) behavior at high temperatures is generic for itinerant ferromagnets only.\cite{T28} Itinerant antiferromagnets will typically show a weak temperature dependence of $\chi$ above $T_N$.\cite{T28} Although itinerant electrons may accidentally mimic the Curie-Weiss-type behavior at high temperatures, it is more plausible to assume that the Fe$_{1-x}$Co$_x$Ga$_3$ solid solution combines features of the itinerant and localized antiferromagnets. 

In fact, both itinerant and localized magnetism can be envisaged for this system. Metallic conductivity of the Fe$_{1-x}$Co$_x$Ga$_3$ solid solutions with $x\geq 0.075$ implies the sizable concentration of itinerant electrons. On the other hand, the Fe/Co disorder on the transition-metal site may lead to at least partial localization, because the Fe and Co $3d$ states provide dominant contribution at the Fermi level.\cite{T9, T21} The latter mechanism is not operative in FeGa$_{3-y}$Ge$_y$, where the disorder is introduced on the Ga site. Indeed, the Ge containing solid solutions are ferromagnetic, \cite{T15} as predicted by LDA. They do not show any signatures of the localized magnetism. Their Curie-Weiss behavior at high temperatures can be well ascribed to the itinerant ferromagnetism. The ratio of the effective and ordered moments in the Rhodes-Wohlfarth plot follows the trend typical for itinerant ferromagnets (see Fig.~6 in Ref.~\onlinecite{T15}).

Remarkably, the apparent antiferromagnetism of Fe$_{1-x}$Co$_x$Ga$_3$ is not matched by the LDA results that predict the ferromagnetic nature of electron-doped FeGa$_3$ irrespective of the doping mechanism. While band structure calculations do not fully account for the Fe/Co disorder, our results for the ordered supercells suggest that the ferromagnetic nature of Fe$_{1-x}$Co$_x$Ga$_3$ is quite robust within LDA and only weakly depends on the local order of the Fe and Co atoms. Therefore, the discrepancy between LDA and the experiment can not be ascribed to a simple mixing of the Fe and Co atoms forming the \mbox{Fe--Fe}, Fe--Co, and Co--Co dumbbells. It rather pertains to a more complex interplay of the itinerant and localized electrons that arise from this mixing. This conjecture is in line with the computational results by Singh,\cite{T21} who was able to stabilize an antiferromagnetic solution in pure FeGa$_3$, but only by adding the on-site Coulomb repulsion $U$ that creates local moments on Fe.

We have shown that in the parent compound FeGa$_3$ the $^{69,71}$Ga spin-lattice relaxation rate $1/T_1(T)$ reveals an unexpected huge maximum at low temperatures with an essentially magnetic relaxation mechanism indicating the presence of an enhanced density of in-gap states placed near the Fermi energy. These states frequently are assigned being responsible for the giant thermopower in Fe based semimetals at low temperatures.\cite{T7,T11,T29,T30} Only above $\sim\!\!70$\,K, when the in-gap level is completely empty, the nuclear spin-lattice relaxation exhibits a crossover to a phonon mechanism characteristic of quadrupolar nuclei in nonmagnetic systems. The other end member, CoGa$_3$, is a band metal. It demonstrates the metallic Korringa behavior of the spin-lattice relaxation with $1/T_1\propto T$. The mixing of these well-understood FeGa$_3$ and CoGa$_3$ compounds triggers strong and unexpected antiferromagnetic spin fluctuations. Indeed, in the intermediate Fe$_{0.5}$Co$_{0.5}$Ga$_3$ compound $1/T_1(T)$ is strongly (by nearly two orders of magnitude) enhanced due to spin fluctuations, with $1/T_1\propto T^{1/2}$ in perfect agreement with Moriya's spin-fluctuation theory for itinerant magnetic systems. Such a $1/T_1(T)$ behavior is a unique feature of weakly and nearly AF metals. The Fe$_{1-x}$Co$_x$Ga$_3$ compounds with $x$ close to 0.5 seem to be very close to magnetic ordering, which is prohibited probably by strong spin fluctuations and the structural disorder between the different T--T dumbbells, in contrast to the FeGa$_{3-y}$Ge$_y$ system, which has a regular arrangement of solely homoatomic Fe--Fe dumbbells and exhibits a FM order at certain doping values. 

\begin{acknowledgments}
This work was supported in part by the MSU priority program; Russian Foundation for Basic Research, grant \# 11-08-00868-a;  joint Russian-Taiwan grant RFBR-NSC \# 12-03-92002-NSC\_ a (101-2923-M-006-001-MY2). A.A.G., N.B. and W.K. acknowledge financial support by the Deutsche Forschungsge-meinschaft (DFG) via TRR 80 (Augsburg-Munich). M.S. acknowledges the U.S. National Science Foundation for the support of research on itinerant magnetism in intermetallic compounds via the NSF-CAREER grant DMR-0955353. The work in Tallinn has been supported by the Mobilitas program of the ESF (grant MTT77). A.A.T. acknowledges insightful discussions with Dr. Christoph Geibel and Dr. Deepa Kasinathan. Dr. S.M. Kazakov is acknowledged for his help with XRD experiments.
\end{acknowledgments}

\end{document}